# Influence of illumination on the quantum lifetime in selectively doped single GaAs quantum wells with short-period AlAs/GaAs superlattice barriers

A. A. Bykov, D. V. Nomokonov, A. V. Goran, I. S. Strygin, I. V. Marchishin, A. K. Bakarov

Rzhanov Institute of Semiconductor Physics, Russian Academy of Sciences, Siberian Branch, Novosibirsk, 630090, Russia

The influence of illumination on a high mobility two-dimensional electron gas with high concentration of charge carriers is studied in selectively doped single GaAs quantum wells with short-period AlAs/GaAs superlattice barriers at a temperature $T = 4.2$ K in magnetic fields $B < 2$ T. It is shown that illumination at low temperatures in the studied heterostructures leads to an increase in the concentration, mobility, and quantum lifetime of electrons. An increase in the quantum lifetime due to illumination of single GaAs quantum wells with modulated superlattice doping is explained by a decrease in the effective concentration of remote ionized donors.

## Introduction

Persistent photoconductivity (PPC), which occurs in selectively doped GaAs/AlGaAs heterostructures at low temperatures ($T$) as the result of visible light illumination, is widely used as a method for changing the concentration ($n_e$), mobility ($\mu$) and quantum lifetime ($\tau_q$) of electrons in such two-dimensional (2D) systems [1-5]. PPC is also used in one-dimensional lateral superlattices based on high mobility selectively doped GaAs/AlGaAs heterostructures [6, 7]. One of the causes of PPC is the change in the charge state of DX centers in doped AlGaAs layers under illumination [8, 9]. PPC is undesirable in high mobility heterostructures intended for the manufacturing of field-effect transistors, as it introduces instability into their performance. One of the ways to suppress PPC is to use short-period AlAs/GaAs superlattices as barriers to single GaAs quantum wells [10]. In this case, the sources of free charge carriers are thin δ-doped GaAs layers located in short-period superlattice barriers in which DX centers do not appear.

Another motivation for remote superlattice doping of single GaAs quantum wells is the fabrication of 2D electronic systems with simultaneously high $n_e$ and $\mu$. In selectively doped GaAs/AlGaAs heterostructures, to suppress the scattering of 2D electron gas on a random potential of ionized donors, the charge transfer region is separated from the doping region by an undoped AlGaAs layer (spacer) [4]. High $\mu$ in such a system is achieved due to a "thick" spacer ($d_S > 50$ nm) with a relatively low concentration $n_e \sim 3 \times 10^{15}$ m$^{-2}$. To implement high mobility 2D electron systems with a "thin" spacer ($d_S < 50$ nm) and high $n_e$, it was proposed in [11] to use short-period AlAs/GaAs superlattices as barriers to single GaAs quantum wells (Fig. 1). In this case, the suppression of scattering by ionized Si donors is achieved not only by separation of the regions of doping and transport, but also by the screening effect of X-electrons localized in AlAs layers [11-13].

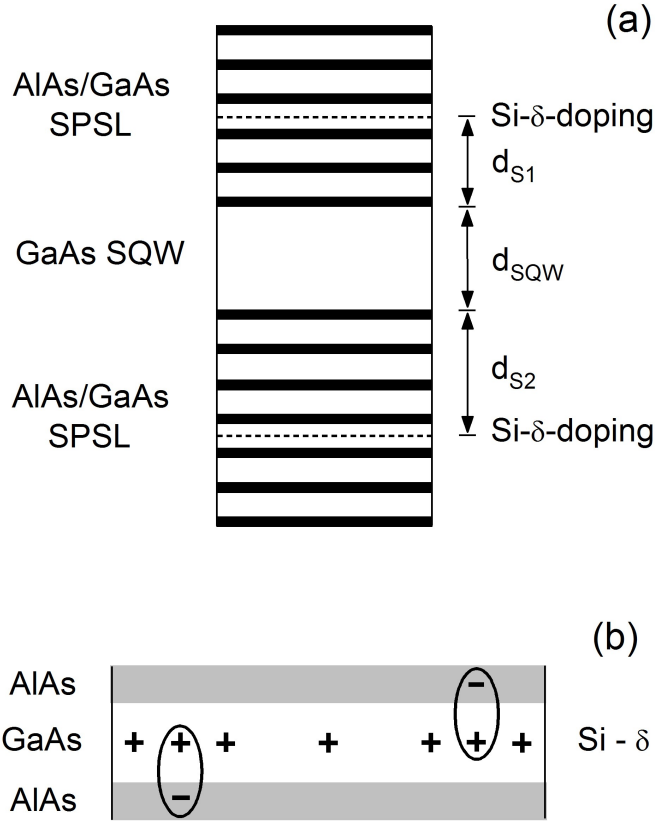

Fig. 1. (a) Schematic view of a single GaAs quantum well with side barriers of short-period AlAs/GaAs superlattices. (b) An enlarged view of a portion of the δ-doped layer in a narrow GaAs quantum well with adjacent AlAs layers. Ellipses show compact dipoles formed by positively charged Si donors in the δ-doped layer and X-electrons in AlAs layers [13].

Superlattice doping of single GaAs quantum wells is used not only to implement high mobility 2D electronic systems with a thin spacer [11, 12], but also to achieve ultrahigh $\mu$ in 2D electronic systems with a thick spacer [14-16]. In GaAs/AlAs heterostructures with modulated superlattice doping, PPC due to a change in the charge states of DX centers should not arise [10]. However, it has been found that in selectively doped single GaAs quantum wells with short-period AlAs/GaAs superlattice barriers and a thin spacer, illumination increases $n_e$ and $\mu$ [17-19], and with a thick spacer, it increases $\tau_q$ [20]. The increase in $\tau_q$ was explained by the redistribution of X-electrons in AlAs layers adjacent to thin δ-doped GaAs layers. However, the effect of illumination on $\tau_q$ in single GaAs quantum wells with a thin spacer and superlattice doping remains unexplored.

One of the features of GaAs/AlAs heterostructures with a thin spacer and superlattice doping grown by molecular beam epitaxy on (001) GaAs substrates is the anisotropy of $\mu$ [21]. In such structures, $\mu_y$ in the [-110] crystallographic direction can exceed $\mu_x$ in the [110] direction by several times [22]. The anisotropy of $\mu$ is due to scattering on the heterointerface roughness oriented along the [-110] direction and arising during the growth of heterostructures [23, 24]. This work is devoted to studying the effect of illumination on a 2D electron gas with an anisotropic $\mu$ in single GaAs quantum wells with a thin spacer and superlattice doping. It has been established that illumination increases $n_e$, $\mu$, and $\tau_q$ in the heterostructures under study. It is shown that the increase in $\tau_q$ after illumination is due to a decrease in the effective concentration of remote ionized donors.

## Quantum lifetime

The traditional method of measuring $\tau_q$ in a 2D electron gas is based on studying the dependence of the amplitude of the Shubnikov – de Haas (SdH) oscillations on the magnetic field ($B$) [25-30]. In 2D electron systems with isotropic $\mu$ low field SdH oscillations are described by the following relation [28]:

$$\rho^{SdH} = 4\, \rho_0\, X(T)\, \exp(-\pi/\omega_c\tau_q)\, \cos(2\pi\varepsilon_F/\hbar\omega_c - \pi), \qquad (1)$$

where $\rho^{SdH}$ is the oscillating component of the dependence $\rho_{xx}(B)$, $\rho_0 = \rho_{xx}(B = 0)$ is the Drude resistance, $X(T) = (2\pi^2 k_B T/\hbar\omega_c)/\sinh(2\pi^2 k_B T/\hbar\omega_c)$, $\omega_c = eB/m^*$, $\varepsilon_F$ is the Fermi energy. Using the results of [26], it is easy to generalize (1) for a 2D system with anisotropic mobility $\mu_d$. In this case, the normalized amplitude of SdH oscillations will be determined by the following expression [31]:

$$A_d^{SdH} = \rho_d^{SdH}/\rho_{0d}\, X(T) = A_{0d}^{SdH} \exp(-\pi/\omega_c\tau_{qd}), \qquad (2)$$

where the index $d$ corresponds to the main mutually perpendicular directions $x$ and $y$, and $A_{0d}^{SdH} = 4$.

The value of $\tau_q$ in single GaAs quantum wells with short-period AlAs/GaAs superlattice barriers is determined mainly by small-angle scattering [11, 12]. In this case, $\tau_q$ can be expressed by the relation [32-34]:

$$\tau_q \cong \tau_{qR} = (2m^*/\pi\hbar)\, (k_F d_R)/n_R^{eff}, \qquad (3)$$

where $\tau_{qR}$ is the quantum lifetime upon scattering on a random potential of a remote impurity, $k_F = (2\pi n_e)^{0.5}$, $d_R = (d_S + d_{SQW}/2)$, $d_{SQW}$ is the thickness of a single GaAs quantum well, and $n^{eff}_R$ is the effective concentration of remote ionized donors. The value of $n^{eff}_R$ takes into account the change in the scattering potential of remote donors when they are bound to X-electrons (Fig. 1b) [13]. The dependence of $n^{eff}_R$ on $n_e$ in the heterostructures under study is described by the following phenomenological relation [35]:

$$n^{eff}_R = n^{eff}_{R0}/\{\exp[(n_e - a)/b] + 1\} \equiv n^{eff}_{R0}\, f_{ab}(n_e), \qquad (4)$$

where $n^{eff}_{R0}$, $a$ and $b$ are fitting parameters. $f_{ab}$ is the fraction of ionized remote donors not associated with X-electrons into compact dipoles.

## Samples under study and details of the experiment

The GaAs/AlAs heterostructures under study were grown using molecular beam epitaxy on semi-insulating GaAs (100) substrates. They were single GaAs quantum wells with short-period AlAs/GaAs superlattice barriers [11, 12]. Two Si δ-doping layers located at distances $d_{S1}$ and $d_{S2}$ from the upper and lower heterointerfaces of the GaAs quantum well served as the sources of electrons. L-shaped bridges oriented along the [110] and [-110] directions were fabricated based on the heterostructures grown by optical lithography and liquid etching. The bridges were 100 µm long and 50 µm wide. The bridge resistance was measured at an alternating current $I_{ac}$ < 1 µA with a frequency $f_{ac}$ ~ 0.5 kHz at a temperature $T$ = 4.2 K in magnetic fields $B$ < 2 T. A red LED was used for illumination.

Table 1. Heterostructure parameters: $d_{SQW}$ is the quantum well thickness; $d_S = (d_{S1} + d_{S2})/2$ is the spacer average thickness; $n_{Si}$ is the total concentration of remote Si donors in δ-doped thin GaAs layers; $n_e$ is the electron concentration; $\mu_x$ is the mobility in the [110] direction; $\mu_y$ is the mobility in the direction [-110]; $\mu_y/\mu_x$ is the mobility ratio. The asterisk marks the values obtained after illumination.

| Structure number | $d_{SQW}$ (nm) | $d_S$ (nm) | $n_{Si}$ ($10^{16}$ m$^{-2}$) | $n_e$ ($10^{15}$ m$^{-2}$) | $\mu_y$ (m$^2$/V s) | $\mu_x$ (m$^2$/V s) | $\mu_y/\mu_x$ |
|---|---|---|---|---|---|---|---|
| 1 | 13 | 29.4 | 3.2 | 7.48 | 124 | 80.5 | 1.54 |
|   |    |      |     | 8.42* | 206* | 103* | 2* |
| 2 | 10 | 10.8 | 5 | 11.5 | 14.7 | 9.33 | 1.58 |
|   |    |      |   | 14.5* | 27.2* | 18.6* | 1.46* |

## Experimental results and discussion

Fig. 2a shows the experimental dependences of $\rho_d(B)$ at $T$ = 4.2 K for heterostructure no. 1 before illumination (curves *1* and *2*) and after illumination (curves *3* and *4*). In the region of $B$ > 0.5 T, SdH oscillations are observed, the period of which in the reverse magnetic field decreased after illumination, which indicates an increase in $n_e$. After illumination, the values of $\rho_{0d}$ also decreased, which is due not only to an increase in $n_e$, but also to an increase in $\mu_d$. The illumination also led to an increase in the positive magnetoresistance (MR) of the 2D electron gas, which indicates an increase in the quantum lifetime [36, 37]. The dependences of $A_d^{SdH}$ on $1/B$ for structure no. 1 are shown in Fig. 2b. In accordance with formula (2), the slope of the dependences $A_d^{SdH}(1/B)$ on a semilogarithmic scale is determined by the value $\tau_{qd}$. A decrease in slope after illumination indicates an increase in $\tau_{qd}$. At the same time, the values of $\tau_{qd}$ measured in the directions [110] and [-110] are equal with an accuracy of 5%.

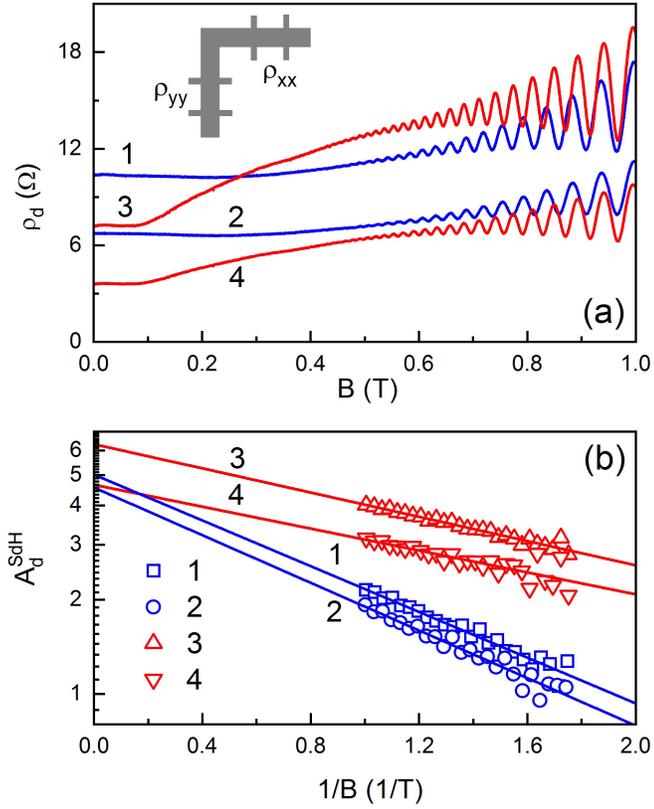

Fig. 2. (a) Experimental dependences of $\rho_d$ on $B$ measured on an L-shaped bridge at $T = 4.2$ K before illumination (*1, 2*) and after illumination (*3, 4*) (no. 1). *1, 3* – $\rho_{xx}(B)$. *2, 4* – $\rho_{yy}(B)$. The inset shows the geometry of the L-shaped bridge. (b) Dependences of $A_d^{SdH}$ on $1/B$ before illumination (*1, 2*) and after illumination (*3, 4*). Symbols are experimental data. Solid lines – calculation by formula (2): *1* – $A_{0x}^{SdH} = 5.02$; $\tau_{qx} = 1.44$ ps; *2* – $A_{0y}^{SdH} = 4.57$; $\tau_{qy} = 1.38$ ps; *3* – $A_{0x}^{SdH} = 6.29$; $\tau_{qx} = 2.72$ ps; *4* – $A_{0y}^{SdH} = 4.66$; $\tau_{qy} = 3.01$ ps.

Fig. 3a shows the experimental dependences of $\rho_d(B)$ at $T = 4.2$ K for heterostructure no. 2 before illumination (curves *1* and *2*) and after illumination (curves *3* and *4*). For this structure, as well as for structure no. 1, short-term illumination at low temperature leads to an increase in $n_e$ and $\mu_d$. However, for structure no. 2, in contrast to no. 1, the dependences $\rho_{xx}(B)$ do not show quantum positive MR, while a classical negative MR is observed [38], which decreases significantly after illumination. Dependences $\tau_{td}(n_e)$ are presented in Fig. 3b. These dependences are not described by the theory [32], which takes into account only the change in $k_F$ with increasing $n_e$, which is due to the change in $n^{eff}_R$ after illumination. A similar behavior of $\tau_{td}$ on $n_e$ is also observed when the concentration of the 2D electron gas is changed using a Schottky gate [12, 35].

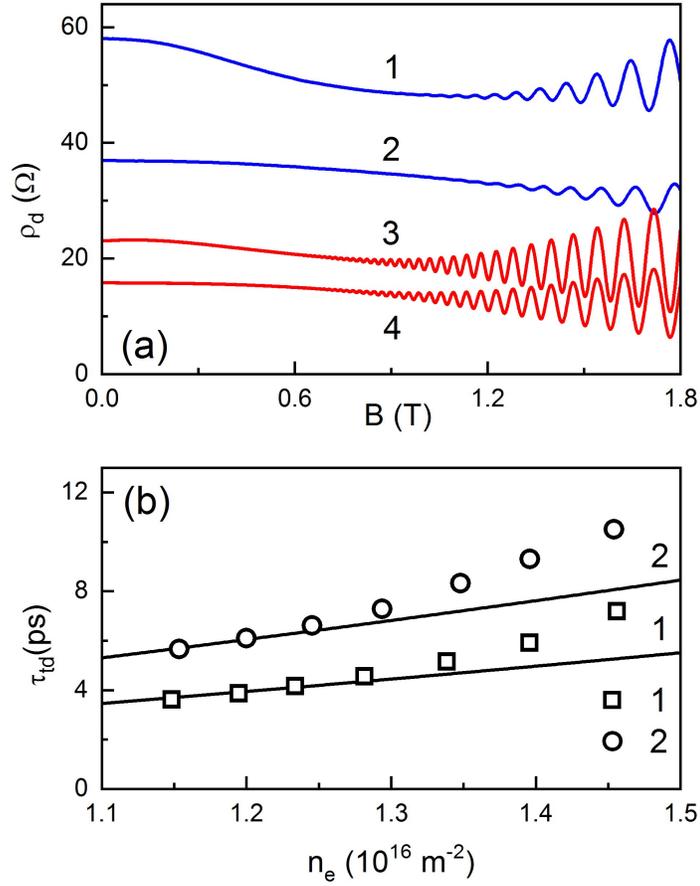

Fig. 3. (a) Dependences of $\rho_{xx}(B)$ and $\rho_{yy}(B)$ measured on the L-shaped bridge at $T = 4.2$ K (no. 2): *1, 2* – before illumination; *3, 4* - after short-term illumination by a red LED. (b) Dependencies of $\tau_{tx}(n_e)$ and $\tau_{ty}(n_e)$. Squares and circles - experimental data: *1* - $\tau_{tx}$; *2* - $\tau_{ty}$. Solid lines – calculation according to the formula: $\tau_{td} \propto n_e^{1.5}$: *1* – $\tau_{tx}$; *2* – $\tau_{ty}$.

The experimental dependences $\tau_{qd}(n_e)$ for structure no. 2 (Fig. 4a) show that $\tau_{qd}$ for different crystallographic directions are equal with an accuracy of 5%, which agrees with [31]. The experimental data are well described by formula (3) for the effective concentration of positively charged Si donors calculated by formula (4). The agreement between the experimental dependences $\tau_{qd}(n_e)$ and the calculated one indicates that the increase in the quantum lifetime of electrons in a single GaAs quantum well after low-temperature illumination is due to a decrease in $n^{eff}_R$.

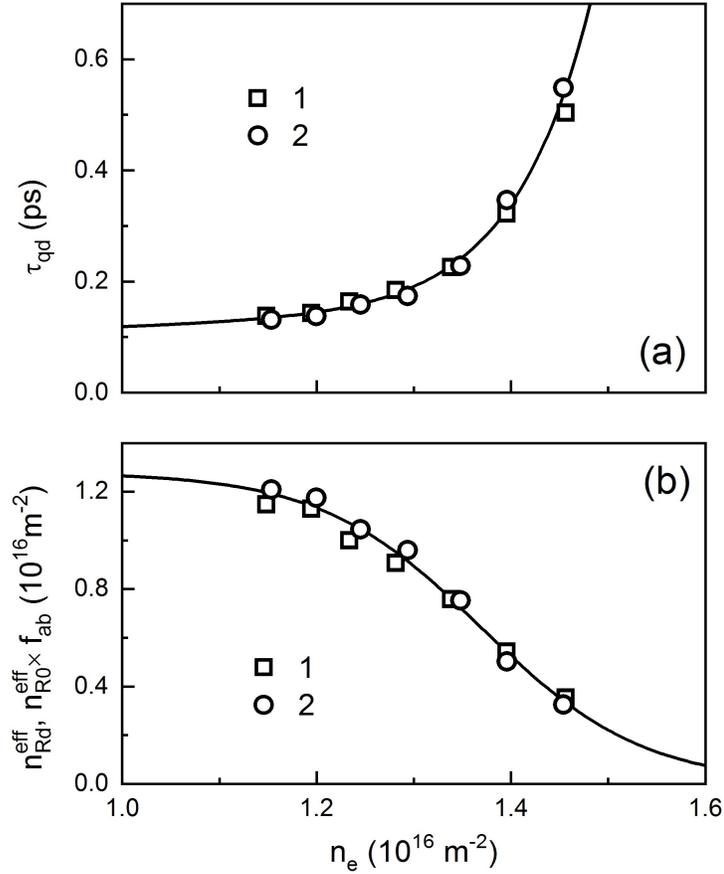

Fig. 4. (a) Dependences of $\tau_{qd}(n_e)$: squares are the experimental values of $\tau_{qy}$; circles – experimental values of $\tau_{qx}$; the solid line is the calculation for $n^{eff}_R = n^{eff}_{R0} f_{ab}$. (b) Dependences of $n^{eff}_R$ and $n^{eff}_{R0} f_{ab}$ on $n_e$: squares and circles are the values of $n^{eff}_R$ calculated from the experimental values of $\tau_{qx}$ and $\tau_{qy}$; solid line – $n^{eff}_{R0} f_{ab}$ for $n^{eff}_{R0} = 1.26 \times 10^{16}$ m$^{-2}$, $a = 1.37 \times 10^{16}$ m$^{-2}$ and $b = 0.082 \times 10^{16}$ m$^{-2}$.

## Conclusion

The influence of illumination on the low-temperature transport in a 2D electron gas with anisotropic mobility in selectively doped single GaAs quantum wells with short-period AlAs/GaAs superlattice barriers in classically strong magnetic fields was studied. It has been shown that, in the heterostructures under study, illumination by a red LED at low temperatures leads to an increase in the concentration, mobility, and quantum lifetime of electrons. An increase in the quantum lifetime of electrons in single GaAs quantum wells with modulated superlattice doping after illumination is explained by a decrease in the effective concentration of remote ionized donors.

**Funding**

This work was supported by the Russian Foundation for Basic Research (project no. 20-02-00309).